# Data Driven Finite Element Method: Theory and Applications

M. Amir Siddiq[a,*]

[a]School of Engineering, University of Aberdeen, Fraser Noble Building, AB24 3UE, Aberdeen, United Kingdom

[*]Corresponding Author: amir.siddiq@abdn.ac.uk

**Abstract**

A data driven finite element method (DDFEM) that accounts for more than two material state variables has been presented in this work. DDFEM framework is motivated from (1,2) and can account for multiple state variables, viz. stresses, strains, strain rates, failure stress, material degradation, and anisotropy which has not been used before. DDFEM is implemented in the context of linear elements of a nonlinear elastic solid. The presented framework can be used for variety of applications by directly using experimental data. This has been demonstrated by using the DDFEM framework to predict deformation, degradation and failure in diverse applications including nanomaterials and biomaterials for the first time. DDFEM capability of predicting unknown and unstructured dataset has also been shown by using Delaunay triangulation strategy for scattered data having no structure or order. The framework is able to capture the strain rate dependent deformation, material anisotropy, material degradation, and failure which has not been presented in the past. The predicted results show a very good agreement between data set taken from literature and DDFEM predictions without requiring to formulate complex constitutive models and avoiding tedious material parameter identification.

## 1. Introduction

Computational modelling of deformation and failure in materials and structures has been under investigation for many decades and still poses a number of challenges. A number of computational mechanics-based modelling techniques exist, such as macroscale finite element methods (3,4), crystal plasticity finite element methods (5–8), boundary element methods (9,10), peridynamics (11,12), finite difference methods (13–15), discrete element methods (16,17), and smoothed particle hydrodynamics (18). An essential part of these modelling technique is material constitutive models. Formulating such material constitutive models pose a number of challenges and is under rigorous research to date. Some of these challenges include, formulation of complex material constitutive models (for e.g. (19,20,29,21–28) and references therein) which incorporate underlying physical mechanisms, and identification of a large number of material parameters (19,24,37,38,27,30–36). Presently, no unified material constitutive model exists which incorporates all physical mechanisms and their interactions. This is due to the complexities associated with the active mechanisms and their interactions. Therefore, depending upon the active mechanisms and length scales different material constitutive models exist. These models can be classified into many different types, for example based on mathematical principles being used (for e.g. variational principles (20)), or to incorporate specific microscale phenomenon (39), or to simulate specific type of manufacturing process (34,40).

Since the creation and development of material databases (41,42), a number of researchers have started to develop data driven (DD) computing in the context of boundary value problems (1,2,43–47) and nonparametric regression approach (48). Such approaches directly use the experimental data and eliminate the efforts, uncertainties and errors induced during inverse modelling to generate stress-strain curves. Kirchdoerfer and Ortiz (1,2) are the pioneers in this regards who introduced a new paradigm of data driven computing by incorporating materials response directly through experimental data. It was demonstrated that the data driven approach merges to classical approach for high fidelity dataset. Nguyen and Keip (45) and Chinesta et al. (44) extended the work of Kirchdoerfer and Ortiz (1) to nonlinear elasticity in the context of finite element methods. A relationship for tangent stiffness matrix and force vectors was derived directly through material data points. Model was then validated to show convergence and accuracy based on the availability of sufficient data points. Leygue et al. (43,49) and Stainier et al. (50) used non-homogeneous strain fields (measured through digital image correlation) of a nonlinear-elastic and elastic-plastic (under monotonous radial loading) materials to build a database of stress-strain couples. Response of truss structures and 2D solids was simulated using the database and data driven computing developed by Kirchdoerfer and Ortiz (1,2). Kirchdoerfer and Ortiz (47) enhanced the robustness of their previous works (1,2) by using a cluster analysis technique for noisy data. It was shown that data driven computing can be enhanced in the context of robustness with respect to outliers by assigning the data points a variable relevance through maximum entropy estimation and distance to the solution dependence.

In the present work, data driven approach presented in (1,2) is extended and implemented in the context of boundary value problem using linear finite element methods. As compared to previous works, this research deals with multiple state variables, i.e. stresses, strains, strain rates, failure stress, material degradation, and anisotropy. Presented DDFEM framework can be applied to a variety of problems in solid mechanics which is also demonstrated.

## 2. Data Driven Finite Element Method

The classical formulation has been discussed elsewhere (for details see ref (1,2,47)) and is not repeated here for brevity. A brief summary of the existing model with emphasis on the extension is discussed in the following. Data driven framework is implemented in the context of linear finite element methods. Starting with the corresponding phase space for three-dimensional boundary value problem, which comprises of set ($\underline{\sigma}, \underline{\varepsilon}, \underline{\dot{\varepsilon}}, \sigma_f, t_d, \underline{\alpha}_{or}$) of stresses, strains, strain rates, failure stress, degradation time, and material anisotropy (orientation angles), respectively. For the three-dimensional problem the corresponding phase space is assumed to be 23 dimensional with $\underline{\sigma}$, $\underline{\varepsilon}$, and $\underline{\dot{\varepsilon}}$ are six dimensional each whereas $\sigma_f$, $t_d$, are scalar and $\underline{\alpha}_{or}$ is a three-dimensional vector of orientation angles. It will be shown later that any combination of these state variables can be used for a specific problem which will decide the exact dimensions of the phase space.

## 2.1 Data Driven Formulation

A discretised finite element model with linear elements of a nonlinear elastic solid is considered as a starting point (for more details please see ref (3)). Each element ($e$) is comprising of $N$ nodes and $M$ gauss points. The discretised model undergoes displacements $\underline{u}$ which is given by $\underline{u} = \xi_a(x, y, z)\underline{u}_a$ with sum on $a$ and $a = 1, \dots, N$. Where $\underline{u}_a$ being nodal displacement due to applied nodal forces $\underline{f}_a$, and $\xi_a$ are the interpolation (shape) functions which are based on linear element and are given by for more details please see ref (3) and therein)

$$\xi_a(x, y, z) = \frac{1}{8}\Sigma^a + \frac{1}{4}x\Lambda_1^a + \frac{1}{4}y\Lambda_2^a + \frac{1}{4}z\Lambda_3^a + \frac{1}{4}yz\Gamma_1^a + \frac{1}{4}xz\Gamma_2^a + \frac{1}{4}xy\Gamma_3^a \frac{1}{4}xyz\Gamma_4^a \qquad \text{Equation 1}$$

with

$\Sigma^a = [+1 \quad +1 \quad +1 \quad +1 \quad +1 \quad +1 \quad +1 \quad +1]$
$\Lambda_1^a = [-1 \quad +1 \quad +1 \quad -1 \quad -1 \quad +1 \quad +1 \quad -1]$
$\Lambda_2^a = [-1 \quad -1 \quad +1 \quad +1 \quad -1 \quad -1 \quad +1 \quad +1]$
$\Lambda_3^a = [-1 \quad -1 \quad -1 \quad -1 \quad +1 \quad +1 \quad +1 \quad +1]$
$\Gamma_1^a = [+1 \quad +1 \quad -1 \quad -1 \quad -1 \quad -1 \quad +1 \quad +1]$
$\Gamma_2^a = [+1 \quad -1 \quad -1 \quad +1 \quad -1 \quad +1 \quad +1 \quad -1]$
$\Gamma_3^a = [+1 \quad -1 \quad +1 \quad -1 \quad +1 \quad -1 \quad +1 \quad -1]$
$\Gamma_4^a = [-1 \quad +1 \quad -1 \quad +1 \quad +1 \quad -1 \quad +1 \quad -1]$

For a known material dataset, i.e. local phase space ($M_e$), data driven framework searches for optimal local state of each element of the material or structure while at the same time satisfying compatibility (eq. 2a) and equilibrium (eq. 2b), viz.

$$\underline{\varepsilon}_e = \sum_{a=1}^{N} \underline{B}_{ea}\underline{u}_a, \text{ and } \sum_{e=1}^{M} w_e \underline{B}_{ea}^T \underline{\sigma}_e = \underline{f}_a \qquad \text{Equation 2(a, b)}$$

where $\underline{\sigma}_e$ is the stress tensor in the element ($e$) while $\underline{B}_{ea}$ is strain matrix and corresponds to the finite element mesh and connectivity.

As mentioned above, material data set ($M_e$) can be comprised of a number of state variables ($\beta_i, i = 1, \dots, n$). For the present work $n = 23$, i.e. stresses ($\underline{\sigma}$), strains ($\underline{\varepsilon}$), and strain rates ($\underline{\dot{\varepsilon}}$), with six components each while scalar failure stress ($\sigma_f$) and degradation time ($t_d$), and finally material anisotropy ($\underline{\alpha}_{or}$) is a three dimensional vector of orientation angles, respectively, are known material states.

For the current multi-state data set, following penalty function $F_e$ is used

$$F_e(\beta_i) = \min_{\beta'_i \in M_e} \sum_{i=1}^{n} C_i (\beta_i - \beta'_i)^2 \qquad \text{Equation 3}$$

with the minimum is searched for all local states in the data set ($M_e$). Here $C_i$ is a numerical value and does not represent a material property while $\beta_i$ are known state variables in material data set ($M_e$) and $\beta'_i$ are unknown state variables to be obtained through minimisation.

Overall objective of the solver is to minimise the global $F$ by enforcing conservation law and compatibility constraints as mentioned above (Equation 2)

$$F = \min_{\beta'_i \in M_e} \sum_{e=1}^{M} w_e F_e(\beta_i) \qquad \text{Equation 4}$$

with $w_e$ being the volume of the element $e$ in undeformed configuration $V_o^e$.

Finally, general form of equilibrium constraint is given by

$$\delta \left( \sum_{e=1}^{M} w_e F_e \left( \sum_{a=1}^{N} \underline{B}_{ea} \underline{u}_a, \underline{\sigma}_e, \underline{\dot{\varepsilon}}_e, \sigma_{fe}, t_{de}, \underline{\alpha}_{or} \right) - \sum_{a=1}^{N} \left( \sum_{e=1}^{M} w_e \underline{B}_{ea}^T \underline{\sigma}_e - \underline{f}_a \right) \eta_a \right) = 0 \quad \text{Equation 5}$$

Following standard procedure of taking possible variations, a system of linear equations is obtained for nodal displacements, the local stresses and the Lagrange multipliers and is given by

$$G^a(\underline{u}_b) = 0, a, b = 1, \ldots, N \qquad \text{Equation 6}$$

Note: For dynamic analyses, inertia can be incorporated separately as

$$M^{ab} \underline{\ddot{u}}_b + G^a(\underline{u}_b) = 0, a, b = 1, \ldots, N \qquad \text{Equation 7}$$

Once all optimal data points are determined, equations 7 are used to define nodal displacements, the local stresses and the Lagrange multipliers.

Note that equations (3) and (4) eliminate the traditional material modelling step which requires material constitutive law, $\underline{\sigma}_e = \hat{\sigma}_e(\underline{\varepsilon}_e, \underline{\dot{\varepsilon}}_e, \sigma_{fe}, t_{de}, \underline{\alpha}_{or})$. Such constitutive models comprise of a large number of unknown material parameters which are required to be identified through tedious inverse modelling (for details see (35,37,51–53) and references therein) and is eliminated through DDFEM framework.

## 3. Results and Discussions

The applicability of the presented formulation is demonstrated by performing simulations on deformation and failure in different applications. All simulations are based on static analyses in the context of finite element methods using equation (6) and linear interpolation (shape) functions. Brief description of individual experiments and simulations from literature along with DDFEM results are presented in the following.

### 3.1 Compressive Mechanical Properties and Degradation of Nanocomposite Bone Scaffolds

Nanocomposite scaffolds are generally used for bone tissue regeneration. The design, selection, structural integrity and changes in mechanical properties of these scaffolds are significantly affected due to degradation and tissue regeneration throughout the process. A detailed review on different types of nanocomposite scaffolds can be found in (54) and references therein. Hydroxyapatite (HAP) is one of the widely used synthetic biomaterial due to its enhanced biocompatibility, and other salient features. However, the HAP is brittle in nature and therefore is used with polycaprolactone (PCL) filler to improve strength. Sharma et al. (54) proposed a multiscale approach to design a three-dimensional PCL /in situ HAP-clay scaffold. A three-dimensional FE model of a scaffold was constructed through micro computed tomography which was then simulated under compression and results were compared with experimental compression test data. Investigations were performed to understand the effect of accelerated degradation on mechanical response of scaffolds. The accelerated degradation studies were performed in alkaline solution (0.1M NaOH). All scaffold samples were placed at 37°C and 5% $CO_2$. Samples were removed after different duration of exposure times, i.e. 1, 5, 7, 14, and 18 days. These samples were then tested under compression to understand the effect of degradation on the stress-strain response of the scaffolds. It was reported that stress-strain response went down as the degradation (exposure) time was increased. Based on these results an analytical model was developed in (54) to predict the scaffolds stress-strain response for measured degradation (exposure) time. Relationship for degradation (54) was obtained through regression as a function of degradation (exposure) time.

As a first part of application of present DDFEM framework, compression test data from Sharma et al. (54) for above mentioned exposure times is used as an input in the model. Compression test simulations are performed with a phase space of 13 dimension with stress ($\boldsymbol{\sigma}$) and strain ($\boldsymbol{\varepsilon}$) being six dimensional, whereas degradation/exposure time ($t_d$) is scalar. A single linear data driven hexahedron finite element is used to simulate the compression tests with three surfaces fixed in orthogonal directions while one surface is compressed. Remaining two surfaces were kept free. Simulations are performed for known degradation times of 1, 14, and 18 days (Figure 1) which showed exactly the same response as experimental data confirming that DDFEM is able to take the experimental data points directly through DDFEM by minimising Equation 5. Since in the above paper (54) an effort was made to come up with a relationship for degradation as a function of degradation days using regression, current DDFEM is used to simulate intermediate degradation times without requiring a relationship. This capability is demonstrated by predicting for the intermediate degradation time of 2, 4, 6, 8, 10, 12 and 16 days (Figure 1). This is achieved by interfacing the DDFEM code with MATLAB and using its built-in interpolation capabilities of scattered data having no structure or order using Delaunay. Results show a logical trend and agreement with experiments, for example samples with 4 days exposure show mechanical response closer to 5 days experimental data and same is true for other exposure times.

Sharma et al. (54) also performed FE modelling of the compression test specimen to validate the proposed analytical model. Simulations were performed for the degradation (exposure) times of 1, 5, 7, 14, and 18 days. Contour plots of deformation profile for different exposure times were presented showing lowest amount of vertical displacement in the model with 1-day exposure time while highest being in 18 days model. Same simulations are repeated here without using any constitutive models and parameters and directly using the experimental data presented in Figure 1. FE analysis is performed by constructing a solid cylinder with 40 mm in diameter and 28.2 mm in height. Compression tests are simulated by putting the sample on a flat bottom surface and applying same pressure value from the top. Compression test simulations are performed using 854 linear data driven hexahedron finite elements. Contour plots of deformation (vertical displacement) are plotted in Figure 2 and show very similar trend as reported in (54), also contour values are very similar to the ones reported in Sharma et al. (54).

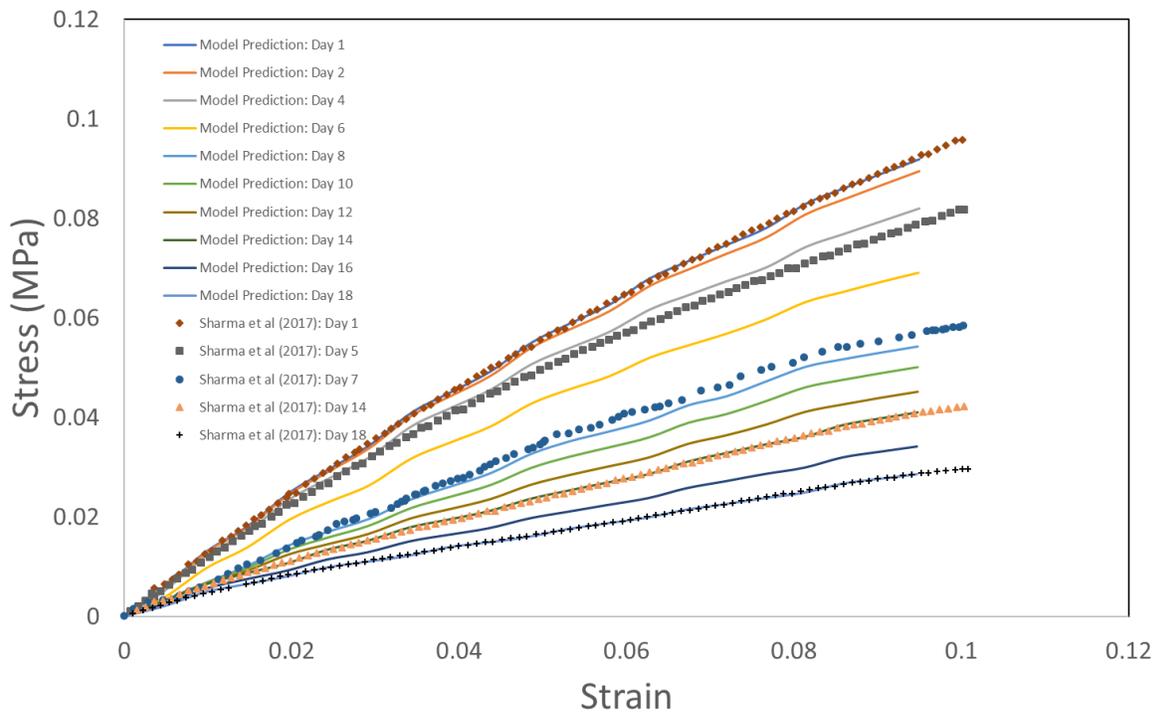

**Figure 1: Compressive stress-strain response of PCL/in situ HAP clay scaffolds: DDFEM vs experiments reported in** (54)

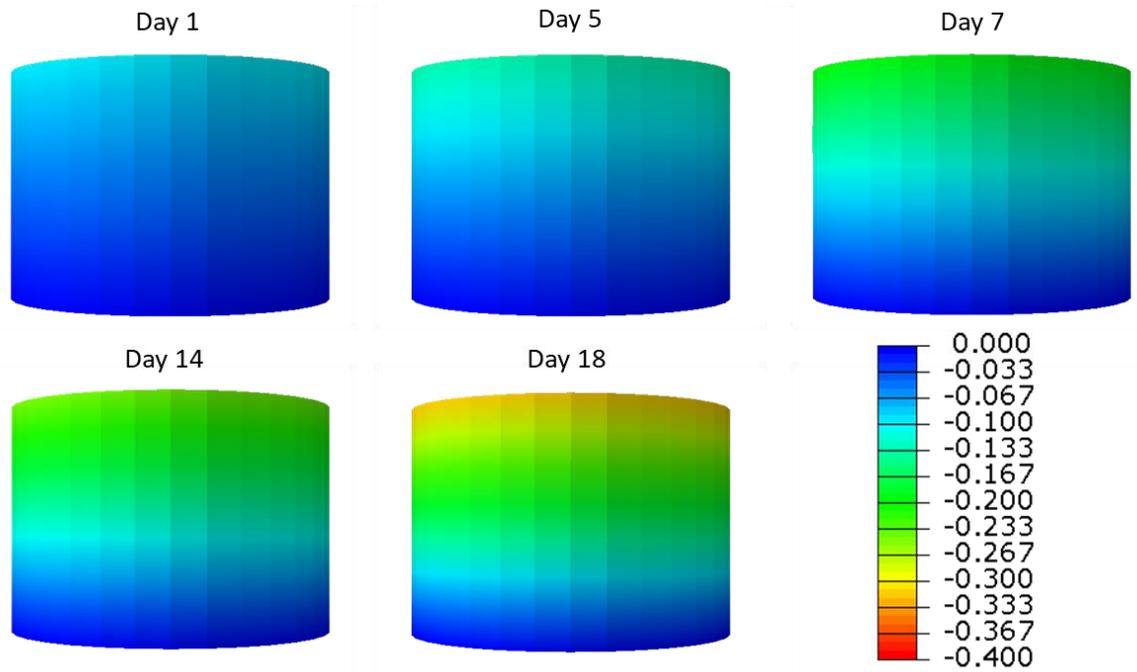

Figure 2: Contour plots of vertical displacement (deformation) in PCL/in situ HAP-clay scaffolds using DDFEM

## 3.2 Tensile Deformation in Oriented Strand Board (OSB) for Wooden Structures

Oriented strand boards (OSB) are manufactured through hot pressing of wood flakes after mixing with wax and adhesive. It is commonly used in wall sheathing, subflooring, roof decking, wood joists, and furniture. Mechanical properties of the OSB panels is highly dependent on the wood strand orientations. Chen and He (55) performed uniaxial tension tests on OSB samples for three different orientations 0°, 45°, and 90°. Tests were conducted in compliance with BS EN 789: 2005. Displacement controlled tests were conducted with a deformation rate of 1.0 mm/min, while to record the final failure phenomenon deformation rate was decreased to 0.5 mm/min once the applied load reached 60% of the maximum load. It was reported that samples showed elastic deformation and failed in a brittle manner with no permanent deformation. It was also reported that stress-strain response was influenced by material orientation with 0° samples showing the stiffest of the response followed by 45° and 90° samples. Based on the tension and similar compression tests, Chen and He (55) proposed a stress-strain relationship involving 6-11 parameters depending on the 1- to 3-dimensional problem.

Present DDFEM framework is used to simulate the tension tests for 3 different OSB orientations, viz. 0°, 45°, and 90°. Tension test simulations were performed with a phase space of 15 dimension with stress ($\underline{\sigma}$) and strain ($\underline{\varepsilon}$) being six dimensional each, whereas material orientation ($\alpha_{or}$) is a three dimensional vector prescribing material orientation with reference to OSB panel major axis (for details see (55)). Results of the tension test data set ((55)) and DDFEM predictions for three different orientations are plotted in Figure 3. Note that experimental data are presented in the form of data clouds. Results show DDFEM capability of directly taking the experimental data and predicting stress-strain response based on local and global equilibrium. Figure 3 also shows that model is able to account for material orientation and predicts the stress-strain response based on material orientations. It can also be inferred from Figure 3 that model is able to identify the suitable data point for each deformation state without any issues after minimising Equations (3-5). It should also be noted that for some deformation values it hits the upper limits while for some it hits intermediate values and for some lower limits. Also, model converges without any difficulty when the next identified data point is below the previous one, i.e. stress reduction. Full scale DDFEM based simulations are also performed for uniaxial tension test. Tension test simulations were performed using 4868 linear data driven hexahedron finite elements. Figure 4 shows the contour plots of von Mises stress in the samples for three orientations (0°, 45°, and 90°) at the same strain value. As expected from Figure 3, 0° orientation shows the stiffest response with highest value of stress while 90° has lowest out of the three orientations. Stress-strain plot of an element from the middle of the sample is also plotted in Figure 4 (right) and it shows stress-strain response is within from individual data sets of three orientations. Also, 0° stress values are higher than 45° and 90°. Figure 5 shows the plot of stress-strain response of two different elements from the 0° specimen. It can be inferred from the Figure 5 that different elements show different stress-strain path based on overall equilibrium of the model, however it must be noted that these local optimised stress-strain points for individual elements are within the data set obtained from experiments.

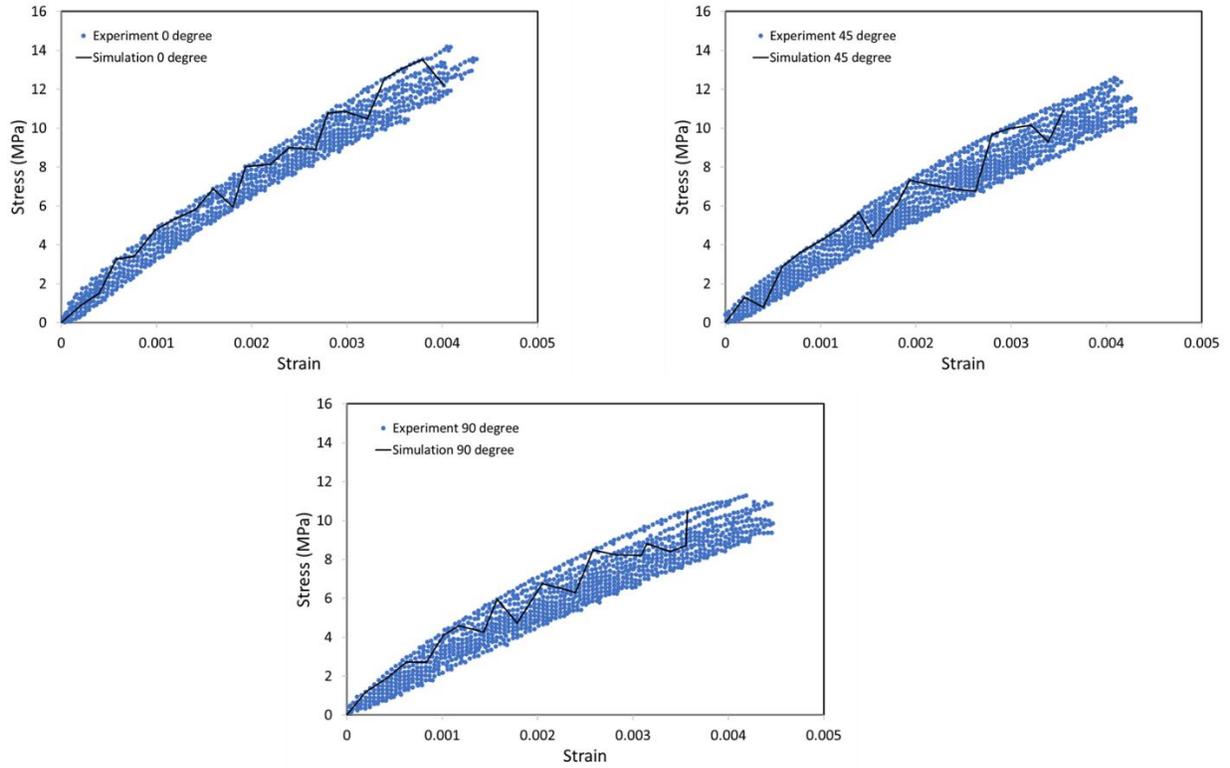

**Figure 3:** Tension test data set for three orientations (0°, 45°, 90°) of OSB and DDFEM predictions

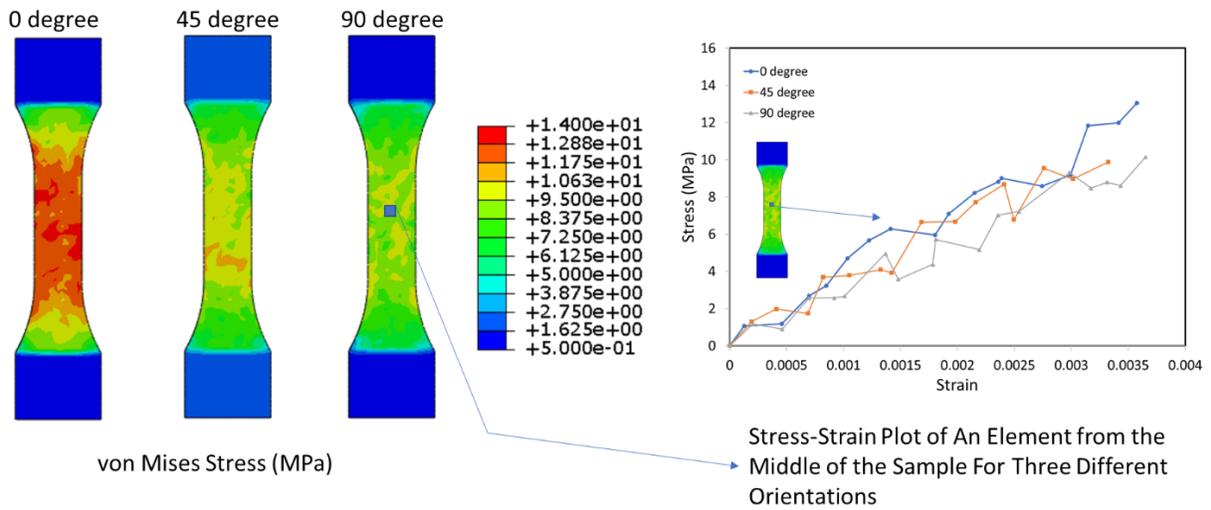

**Figure 4:** DDFEM based simulations of uniaxial tension test performed on test samples based on Chen and He (55)
9

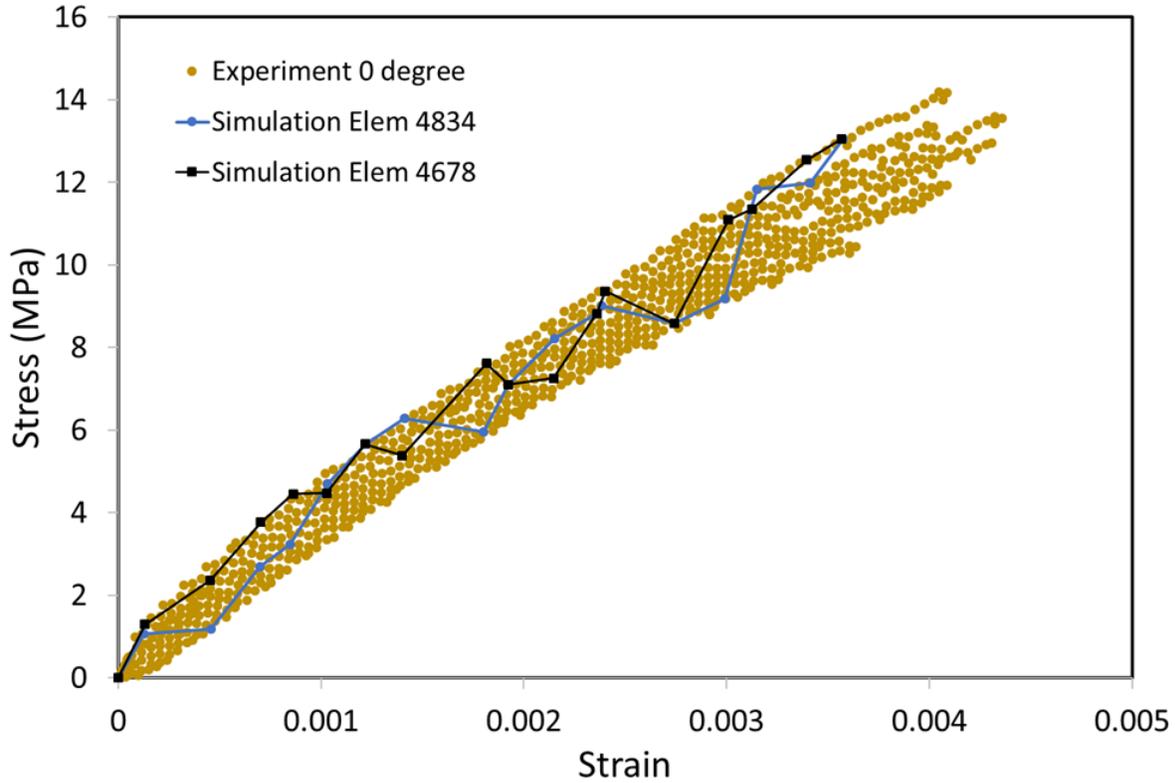

Figure 5: Stress-strain response plot for two different elements in the uniaxial tension test specimen in comparison to the 0° experimental data set

### 3.3 Mechanical Response of Carbon Nanotubes

Kok and Wong (56) performed molecular dynamics (MD) studies on single-walled carbon nanotubes (SWCNT) and double-walled carbon nanotubes (DWCNT) to evaluate their mechanical properties. Mechanical properties were estimated for various aspect ratios and strain rates. For the present study, MD results for (5,5) armchair SWCNT at different strain rates are used. For model application purposes and to check the handling of data size by presented model, data for different strain rates is used to generate the stress strain curves for unknown intermediate strain rates (see surface plot in Figure 6 ) using Delaunay triangulation. At continuum scale, carbon nanotubes have been modelled as shells (57), a beam (58), and a combination of many truss elements (59) (for details and other literature please see references there in). To demonstrate the application of the presented data driven formulation a similar approach has been used, i.e. by modelling the SWCNT as a single truss element under tensile loading. Tensile test simulations are performed with a phase space of 18 dimension with stress ($\underline{\sigma}$), strain ($\underline{\varepsilon}$) and strain rate ($\underline{\dot{\varepsilon}}$) being six dimensional each. Comparison between MD generated stress strain response and model predictions for three different strain rates are shown in Figure 6. Results show a very good agreement between MD simulations and DD model response for multi-state material dataset.

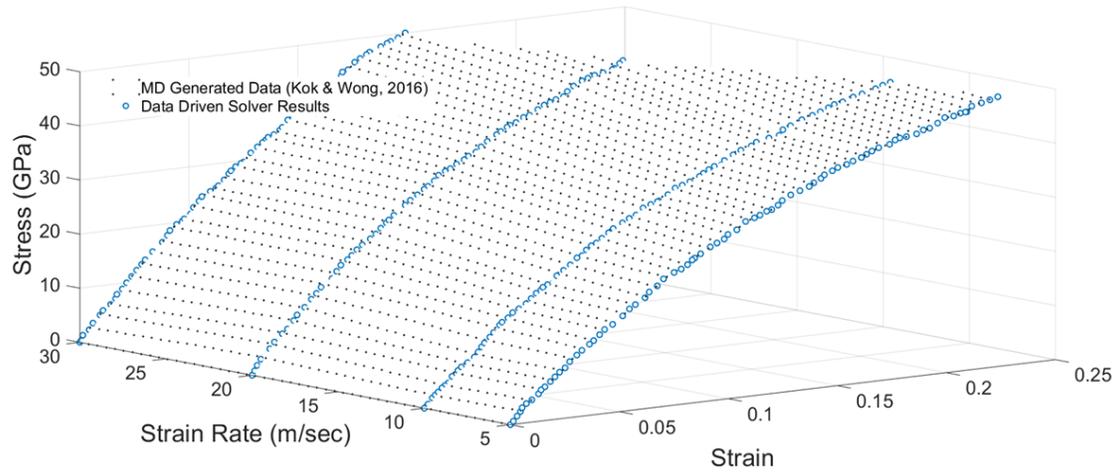

**Figure 6: Comparison of data driven (DDFEM) model prediction and MD generated data of single-walled carbon nanotubes (SWCNT) from Kok and Wong** (56)

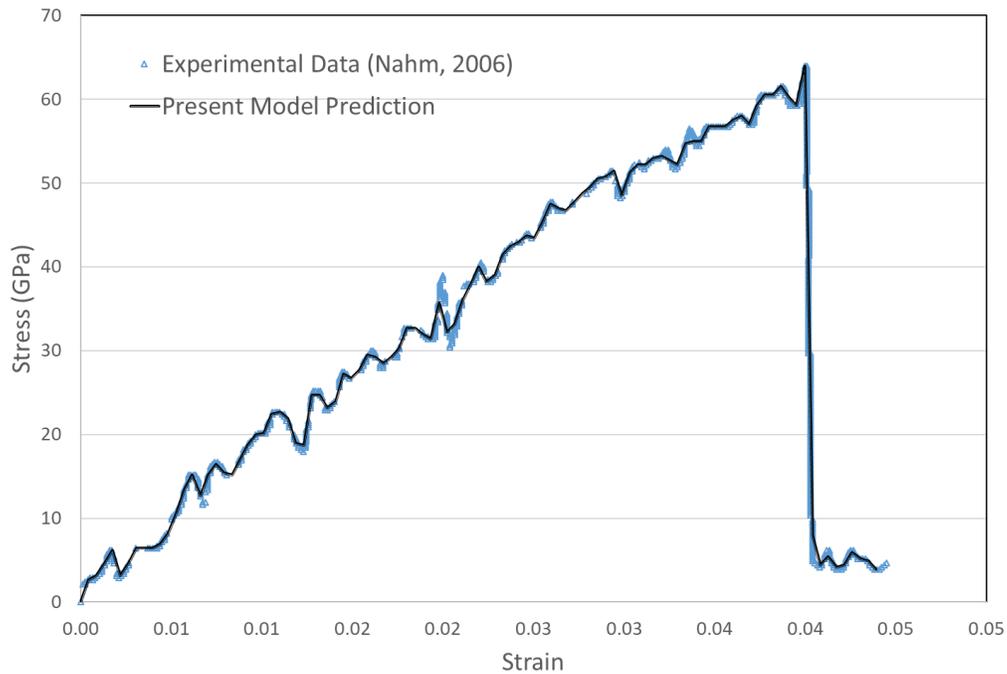

Figure 7: Comparison of DDFEM model prediction and experimental data reported in Nahm (60) **for MWCNT**

Nahm (60) performed tensile testing of multi-walled carbon nanotubes (MWCNT) using a nano-manipulator and sub nano-resolution force sensor in scanning electron microscope (SEM). As explained above, MWCNT was modelled using a truss element. Tensile test simulations are performed with a phase space of 12 dimension with stress ($\underline{\sigma}$), and strain ($\underline{\varepsilon}$) being six dimensional each. The results of the comparison of experimental and predicted stress-strain response are plotted in Figure 7 showing a very good agreement up to the final failure of the nanotube without using any material constitutive model and directly using experimental data.

## 3.4 Mechanical Response of CNT/Epoxy based Nanocomposites

Yu and Chang (61) performed experimental studies to understand tensile behaviour of MWCNT-reinforced epoxy-matrix composites. Effect of the weight fractions and diameters of CNT's on stress-strain behaviour, and strength was investigated. For the model application purposes, CNT/epoxy composite with 1% MWCNT weight data is used. Uniaxial tension test is simulated using three-dimensional eight node cube element with one integration (gauss) point. Tensile test simulations are performed with a phase space of 12 dimension with stress ($\sigma$), and strain ($\varepsilon$) being six dimensional each. A comparison between model predictions and experimental results are presented in Figure 8 showing a good agreement. Note that experimental data are presented in the form of data clouds.

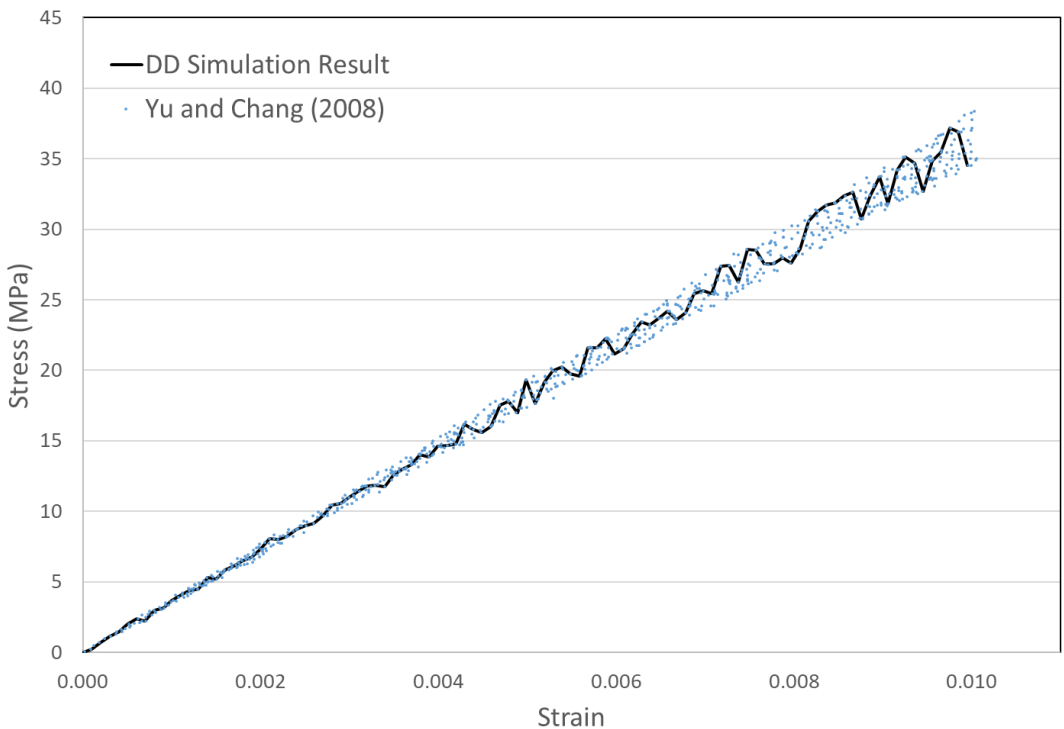

**Figure 8:** Comparison of data driven (DDFEM) model prediction and experimental data of CNT/epoxy composite (1% MWCNT weight fraction) reported in **Yu and Chang** (61)

Finally, in order to demonstrate the application of the proposed model to finite element-based analysis and to show the models ability to capture realistic fracture patterns during deformation; a finite element model of dog-bone sample is used. The sample is based on ASTM D638 which was used by Yu and Chang (61) in above presented example. Experimental data from Figure 8 is directly used without using any material constitutive model. Model was discretised using 4488 reduced integration eight node hex elements. Displacement boundary condition of 1.0mm/min was prescribed which was used during experiments. Contour plots of the von Mises stress obtained from DDFEM and conventional FE solver at two stages of the deformation (pre and post failure) are plotted in Figure 9. For the present study, fracture in the material is modelled by specifying the failure stress value of 32MPa for both conventional and DDFEM simulations. Element was deleted from the simulation once the failure stress is reached. Results show the presented DDFEM's capability of predicting the stresses and macroscopic fracture without any numerical difficulties. Also, results are similar to conventional FE solver in terms of stress distribution and fracture profile. It can be inferred from Figure 9 that overall stresses induced are very similar for both solvers. However, there are slight differences in the stress distribution of pre-failed samples. This is due to the fact that DDFEM identifies different stress values through data cloud while conventional FE solver used elastic constant, i.e. a single stress-strain line. Similarly, for post-failure sample, the stress distribution near the failure zone is different for both solvers. This is due to the fact that DDFEM solver deletes the element after failure and continues the simulation, hence stresses start to reduce (unloading) in the sample. While conventional FE solver used does not have the capability to delete the element and continue the simulation simultaneously, so it stops as the failure stress is reached without deleting the elements, hence elements near the failure don't relax (unloading).

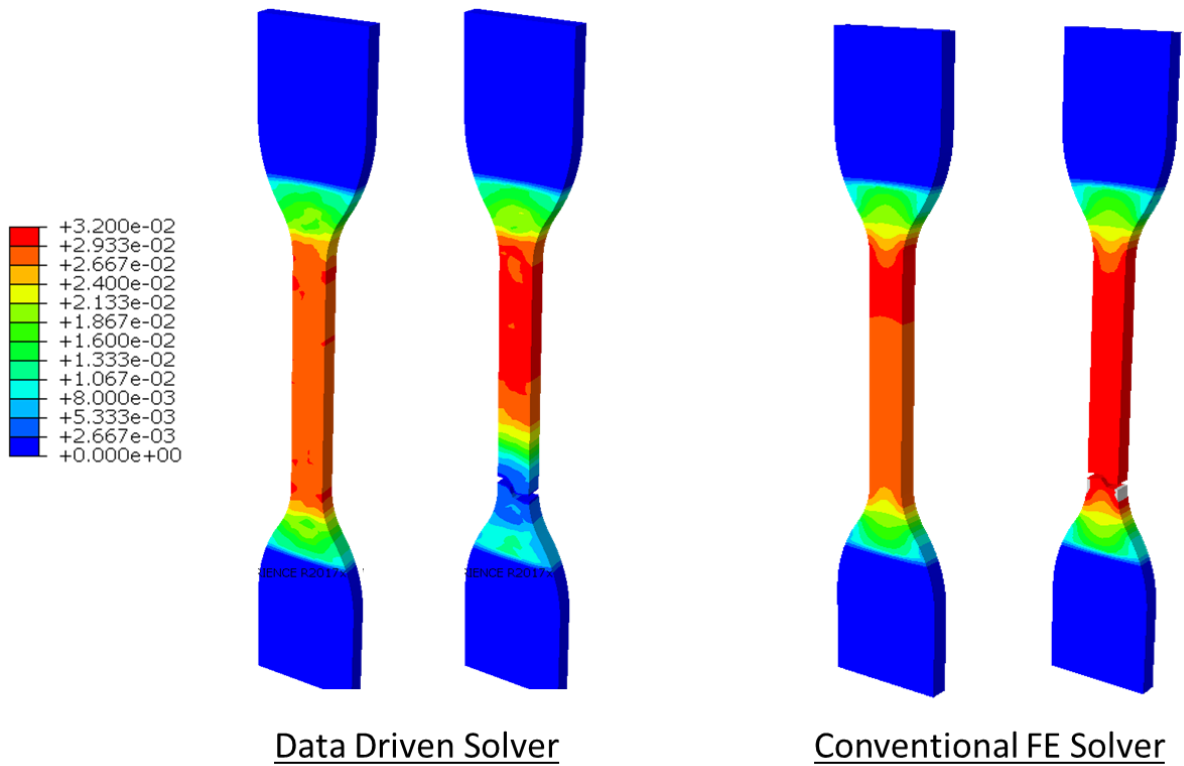

**Figure 9: Comparison of DDFEM predictions with conventional FE solver of deformation and failure in CNT/Epoxy composite (1% wt.)**

## 4. Conclusions

A data driven parameter free finite element method to predict the mechanical response for nanomaterials and biomaterials was presented. DDFEM framework was used to estimate deformation and failure in bone scaffolds, oriented strand board, carbon nanotubes, and nanocomposites by incorporating the interdependencies of stresses, strains, strain rates, failure stress, material degradation, and anisotropy. Numerical predictions were presented based on data driven computing approach and showed a good agreement with the data set taken from experiments and molecular dynamics simulations. The presented model was applied in the context of nano and biomaterials; however, this model can be applied to any length scale with elastic unloading. As a future work, the model will be further extended to account for macro- and micro- plasticity in the context of von Mises theory and microscale slip based plasticity in single crystals. As discussed in section 1, current macro- and micro-scale plasticity models require a large number of parameters which can be avoided if material data set can directly be used to account for plasticity which is under development and will be reported in near future.

## Acknowledgements

No external funding was received for this project.